\documentclass[aps,prc,twocolumn,groupedaddress,showpacs,showkeys,amsmath,amssymb,floatfix,superscriptaddress]{revtex4}
\usepackage{graphicx}% Include figure files
\usepackage{dcolumn}% Align table columns on decimal point
\usepackage{bm}% bold math
\usepackage{graphicx}
\usepackage{times}

\begin{document}

%--------------------------------- frontmatter ---------------------------------

\title{CNO and \emph{pep} neutrino spectroscopy in Borexino: Measurement  of the deep underground
production of cosmogenic $^{11}$C in  organic liquid scintillator}

%\corauth[cor]{E-mail addresses: davide.franco@mi.infn.it, davide.dangelo@lngs.infn.it}

\newcommand{\lngs}{\affiliation{I.N.F.N Laboratori Nazionali del Gran Sasso, SS 17 bis Km 18+910, I-67010 Assergi(AQ), Italy}}
\newcommand{\milano}{\affiliation{Dipartimento di Fisica Universit\`a and I.N.F.N., Milano,
Via Celoria, 16 I-20133 Milano, Italy}}
\newcommand{\princeton}{\affiliation{Dept. of Physics,Princeton University, Jadwin
Hall, Washington Rd, Princeton NJ 08544-0708, USA}}
\newcommand{\tum}{\affiliation{Technische Universit\"at M\"unchen, James Franck Strasse, E15 D-85747, Garching, Germany}}
\newcommand{\college}{\affiliation{Astroparticule et Cosmologie APC, Coll\`ege de France, 11 place Marcelin
Berthelot, 75231 Paris Cedex 05, France}}
\newcommand{\dubna}{\affiliation{Joint Institute for Nuclear Research, 141980 Dubna, Russia}}
\newcommand{\kurchatov}{\affiliation{RRC Kurchatov Institute, Kurchatov Sq.1, 123182 Moscow, Russia}}
\newcommand{\queen}{\affiliation{Queen's University, Physics Department, Kingston, Ontario, Canada K7L 3N6}}
\newcommand{\mpi}{\affiliation{Max-Planck-Institut fuer Kernphysik,Postfach 103 980 D-69029, Heidelberg, Germany}}
\newcommand{\kiev}{\affiliation{Kiev Institute for Nuclear Research, 29 Prospekt Nauki 06380 Kiev, Ukraine}}
\newcommand{\genova}{\affiliation{Dipartimento di Fisica Universit\`a and I.N.F.N., Genova, Via Dodecaneso,33 I-16146 Genova, Italy}}
\newcommand{\perugia}{\affiliation{Dipartimento di Chimica Universit\`a, Perugia, Via Elce di Sotto, 8 I-06123, Perugia, Italy}}
\newcommand{\virginia}{\affiliation{Physics Department, Virginia Polytechnic Institute and State University,
Robeson Hall, Blacksburg, VA 24061-0435, USA}}
\newcommand{\cracovia}{\affiliation{M.Smoluchowski Institute of Physics, Jagellonian University, PL-30059
Krakow, Poland}}
\newcommand{\cleveland}{\affiliation{Case Western Reserve University, Cleveland OH 44118, USA}}
\newcommand{\df}{\email{Davide.Franco@mi.infn.it}}
\newcommand{\dd}{\email{Davide.Dangelo@lngs.infn.it}}
\newcommand{\jm}{\altaffiliation{Present address: La\-bo\-ra\-t\'{o}\-rio de
Ins\-tru\-menta\-\c{c}\~{a}o e F\'{\i}\-si\-ca Ex\-pe\-ri\-men\-tal de
Par\-t\'{\i}\-cu\-las (LIP), Av. Elias Garcia, 14,
1$^{\circ}$, 1000-149 Lisboa, Portugal.}}
\newcommand{\hb}{\altaffiliation{Present address: North Carolina State University, 890 Oval Drive, 
Campus Box 8206, Raleigh, NC 27695-8206, USA}}

\author{H.~Back}\hb\virginia
\author{M.~Balata}\lngs
\author{G.~Bellini}\milano
\author{J.~Benziger}\princeton
\author{S.~Bonetti}\milano
\author{B.~Caccianiga}\milano
\author{F.~Calaprice}\princeton
\author{D.~D'Angelo}\dd\tum
\author{A.~de~Bellefon}\college
\author{H.~de~Kerret}\college
\author{A.~Derbin}\dubna
\author{A.~Etenko}\kurchatov
\author{R.~Ford}\princeton
\author{D.~Franco}\df\milano\mpi
\author{C.~Galbiati}\princeton
\author{S.~Gazzana}\lngs
\author{M.~Giammarchi}\milano
\author{A.~Goretti}\princeton
\author{C.~Grieb}\virginia
\author{E.~Harding}\princeton
\author{G.~Heusser}\mpi
\author{A.~Ianni}\lngs
\author{A.~M.~Ianni}\princeton
\author{V. V. ~Kobychev}\kiev 
\author{G.~Korga}\milano
\author{Y.~Kozlov}\kurchatov
\author{D.~Kryn}\college
\author{M.~Laubenstein}\lngs
\author{C.~Lendvai}\tum
\author{M.~Leung}\princeton
\author{E.~Litvinovich}\kurchatov
\author{P.~Lombardi}\milano
\author{I.~Machulin}\kurchatov
\author{J.~Maneira}\jm\milano\queen
\author{D.~Manuzio}\genova
\author{G.~Manuzio}\genova
\author{F.~Masetti}\perugia
\author{U.~Mazzucato}\perugia
\author{K.~McCarty}\princeton
\author{E.~Meroni}\milano
\author{L.~Miramonti}\milano
\author{M.~E.~Monzani}\lngs
\author{V.~Muratova}\dubna
\author{L.~Niedermeier}\tum
\author{L.~Oberauer}\tum
\author{M.~Obolensky}\college
\author{F.~Ortica}\perugia
\author{M.~Pallavicini}\genova
\author{L.~Papp}\milano
\author{L.~Perasso}\milano
\author{A.~Pocar}\princeton
\author{R.~S.~Raghavan}\virginia
\author{G.~Ranucci}\milano
\author{A.~Razeto}\lngs
\author{A.~Sabelnikov}\lngs
\author{C.~Salvo}\genova
\author{S.~Schoenert}\mpi
\author{T.~Shutt}\cleveland
\author{H.~Simgen}\mpi
\author{M.~Skorokhvatov}\kurchatov
\author{O.~Smirnov}\dubna
\author{A.~Sotnikov}\dubna
\author{S.~Sukhotin}\kurchatov
\author{Y.~Suvorov}\lngs
\author{V.~Tarasenkov}\kurchatov
\author{R.~Tartaglia}\lngs
\author{D.~Vignaud}\college
\author{R. B.~Vogelaar}\virginia
\author{F.~Von~Feilitzsch}\tum
\author{V.~Vyrodov}\kurchatov
\author{M.~W\'ojcik}\cracovia
\author{O.~Zaimidoroga}\dubna
\author{G.~Zuzel}\mpi

\date{\today}

\begin{abstract}

Borexino is an experiment for low energy neutrino spectroscopy at the Gran Sasso underground laboratories. It is designed to measure the mono-energetic $^7$Be solar neutrino flux in real time,
via neutrino-electron
elastic scattering in ultra-pure organic liquid scintillator.
Borexino has the potential to also detect neutrinos from the \emph{pep} fusion process and the
CNO cycle.
For this measurement to be possible, radioactive contamination in the detector must be kept extremely low. Once sufficiently clean conditions are met, the main background source is $^{11}$C, produced
in reactions induced by the residual cosmic muon flux on $^{12}$C. In the process, a free neutron is almost always produced.
$^{11}$C can be tagged on an event by event basis by looking at the three-fold coincidence
with the parent muon track and the subsequent neutron capture on protons.
This coincidence method has been implemented on the Borexino Counting Test Facility data.
We report on the first event by event identification of \emph{in situ} muon induced $^{11}$C
in a large underground scintillator detector.
We measure a $^{11}$C production rate of 0.130 $\pm$ 0.026 (stat) $\pm$ 0.014 (syst) day$^{-1}$ ton$^{-1}$,
in agreement with predictions from both experimental studies performed with a muon beam on a scintillator target and \emph{ab initio} estimations based on the $^{11}$C producing nuclear reactions.
\end{abstract}

\maketitle

%--------------------------------- mainmatter ---------------------------------
\section{Introduction}
\label{sec:intro}

\begin{figure}[ht]
\centering
\resizebox{0.95\columnwidth}{!}{\includegraphics{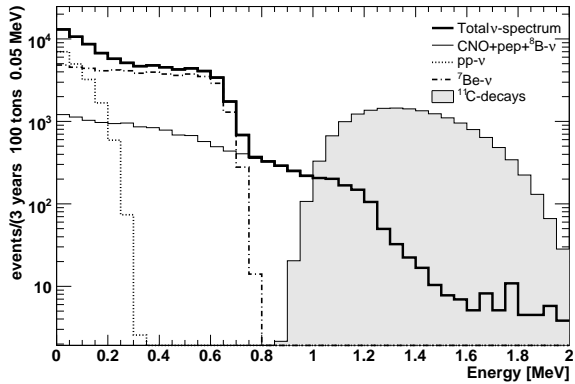}}
\caption{\label{fig:allnu} \linespread{0.95} \small \em
Expected recoil electron energy for different solar neutrinos interacting in Borexino assuming 3 year live time exposure, 
100 tons fiducial volume and a detector energy resolution of 5\%/$\sqrt{E_{[MeV]}}$. Neutrino fluxes are derived assuming the 
 Standard Solar Model BP2004+LUNA \cite{Bah04A,For04A} and the LMA oscillation scenario \cite{Bah03A}. 
The shaded superimposed area is the expected $^{11}C$ background \cite{Hag00A}.}
\end{figure}

Results from solar neutrino \cite{Homestake}
and reactor \cite{KAMLAND}
antineutrino experiments provide compelling evidence
for neutrino oscillations as the explanation of the long-standing solar neutrino problem \cite{Bah92}.
The next goal in solar neutrino physics is probing in real time the low energy ($<$ 2 MeV) component of the solar neutrino spectrum,
which accounts for more than 99\% of the total flux.
This includes neutrinos produced in the
\emph{pp}, $^7$Be, and \emph{pep} nuclear fusion reactions and the CNO-cycle. \\
Particularly, \emph{pep} and CNO neutrinos are an ideal source for probing the energy region, 
between 1 and 3 MeV, at which the transition between matter and vacuum dominated oscillations is 
supposed to occur, according to the MSW-LMA oscillation solution \cite{Bah05A}.
Furthermore, the \emph{pep} and \emph{pp} solar neutrino rates are directly related, via the ratio of the cross section of the two reactions.
Measuring the \emph{pep} neutrino flux is hence a way to study the fundamental \emph{pp}
fusion reaction by which the Sun burns, and improves our knowledge of the solar neutrino
luminosity, thence yielding a crucial check of the Sun stability over a time scale of $10^5 - 10^6$ 
years by comparison with the photon luminosity.
CNO neutrinos play a key role on the age estimation of the Globular Clusters  \cite{For05},
pivotal in setting a lower limit for the age of the universe.\\
Deep underground organic liquid scintillator detectors, like Borexino and KamLAND,
are well positioned to measure \emph{pep} and CNO solar neutrinos. The 1.4 MeV, mono-energetic \emph{pep} neutrinos are particularly well
identifiable by the characteristic Compton-like electron recoil spectrum they produce.
The main challenge they face is the identification and suppression of the $^{11}$C background.
$^{11}$C is produced deep underground by residual cosmic muons interacting with $^{12}$C atoms in the scintillator. The rate of the process is a function of the location and depth of the experiment.
As can be seen in fig. \ref{fig:allnu}, the $^{11}$C background at Gran Sasso falls in the energy region for the detection of \emph{pep} and CNO neutrinos.
In 1996, Deutsch  \cite{Deu96P} suggested that $^{11}$C decays could be detected and subtracted
exploiting the neutron emission in the reaction:
\begin{equation}
\label{eq:muon}
\mu  \; (+ \; secondaries) + ^{12}C  \rightarrow \mu \; (+ \; secondaries) + ^{11}C + n.
\end{equation}
He proposed using a three-fold coincidence which links the parent muon, the neutron capture on protons, and the $^{11}$C decay.
The validity of such technique was studied in detail in \cite{Gal05A}.
We apply the three-fold coincidence technique to data from the Borexino
Counting Test Facility (CTF). This is, to the best of our knowledge, the first \emph{in situ} event by event detection of $^{11}$C production deep underground. We then use our results to
evaluate \emph{pep} solar neutrino detection with Borexino.

\section{$^{11}$C \emph{in situ} production: the three-fold coincidence technique}

$^{11}$C  $\beta^+$-decays  with a mean life of 29.4 min and an endpoint energy of 0.96 MeV:
\begin{equation}
\label{eq:c11}
^{11}C  \rightarrow  ^{11}B + e^+ + \nu_e.
\end{equation}
The total energy released in the detector by the decay and the following positron annihilation is
between 1.02 and 1.98 MeV, partially covering  the best window for the observation of
the \emph{pep}+CNO signal (0.8-1.3 MeV).\\
The probability to produce $^{11}$C nuclides
in muon-induced cascades was experimentally
determined with a target experiment (NA54) on a muon beam at CERN \cite{Hag00A}. The inferred $^{11}$C rate for Borexino and CTF is 0.146 $\pm$ 0.015 day$^{-1}$ ton$^{-1}$
(0.074 $\pm$ 0.008 day$^{-1}$ ton$^{-1}$ in the \emph{pep}+CNO neutrino window). \\
The study reported in \cite{Gal05A} identified  eight different processes for the $^{11}$C production in muon showers
and provided a quantitative estimate for the rate in all the production channels.
The result seems robust in view of the fact that the calculated production rate matches the rate measured at the NA54 CERN facility.\\
Two of the production channels identified,  $^{12}$C(p,d)$^{11}$C and $^{12}$C($\pi^{+}$, $\pi^{0}$+p)$^{11}$C,
do not produce a free neutron in the final state, and therefore escape any possibility of detection
by the three-fold coincidence technique.
These two production channels are referred to as "invisible channels", and they account for 5\% of the
$^{11}$C production rate \cite{Gal05A}.\\
Neutrons are  captured on hydrogen with a capture mean time of $\sim$ 250 $\mu$s in pseudocumene
emitting  a characteristic $\gamma$ of 2.2 MeV. Neutrons can also be captured
on carbon isotopes emitting $\gamma$ with larger energy, but the cross section
is two orders of magnitude lower than on hydrogen.\\
In order to identify and suppress the $^{11}$C background,
each 2.2 MeV  $\gamma$ produced in the scintillator from the muon-induced showers must be localized in space and time.\\
After each  muon-induced neutron detection, the three-fold coincidence technique defines a set
of potential $^{11}$C candidates  within a time delay \emph{t} from the detected muon and
inside a sphere of radius \emph{r} from the neutron capture point.
We assume that no convective currents move the $^{11}$C nuclide from the production point in the time scale
of the $^{11}$C mean life.\\
In Borexino, the $^{11}$C candidates will be discarded  in order to increase
the \emph{pep}+CNO signal to background ratio.
The success of the \emph{pep} and CNO neutrino measurement will depend on two main conditions: the minimization of the
detector  mass-time fraction lost to the cuts implementing the three-fold coincidence  and the achievement of a  high efficiency  in the $^{11}$C suppression.\\
The limited size of CTF represents a challenging test for the three-fold coincidence technique. The goal in CTF
is the measurement of the  $^{11}$C production rate by looking at the time profile of  the $^{11}$C candidates.

\section{Experimental setup}
\label{sec:ctf}

\begin{figure}[t]
\begin{center}
\includegraphics[width=0.95\columnwidth]{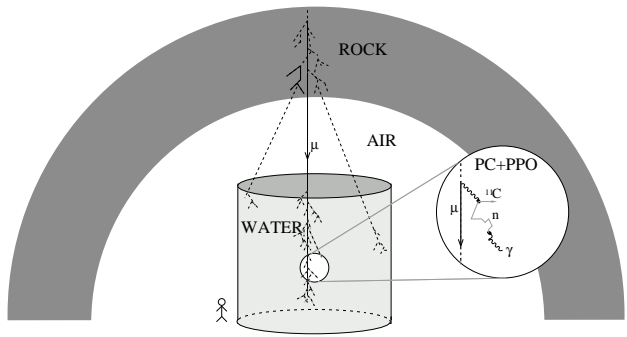}
\caption{\label{fig:sketch}  Overview of the CTF detector and of the physical
processes included in the simulation.}
\end{center}
\end{figure}

CTF \cite{Bor98A} is the Borexino prototype detector installed at the Gran Sasso underground laboratory.
It was designed to test the required radiopurity of the Borexino liquid scintillator and its purification strategy.
The CTF of Borexino was the first detector to prove the level of purities needed for solar neutrino
physics on a multiton-scale, in its 1994-95 campaign \cite{Bor98AB}.
The active detector consists of 3.73 tons (0.88 ton/m$^3$ density) of the Borexino-like scintillator, a mixture of  pseudocumene
(PC, 1,2,4-trimethylbenzene, C$_6$H$_3$(CH$_3$)$_3$) plus 1.5 g/l of 
PPO (2,5-diphenyloxazole, C$_{15}$H$_{11}$NO),  housed in a 1 m radius transparent nylon vessel.
A 7 m diameter stainless steel open structure supports 100 8'' photomultiplier tubes (PMT) equipped with light concentrators
which provide an optical coverage of 21\%.\\
The detector is housed within a cylindrical tank (11 m diameter and 10 m height) containing
1000 tons of pure water, which provides 4.5 m shielding against neutrons from the rock and
external $\gamma$-rays from the rock and from the same PMTs.
16 upward-looking  PMTs mounted on the bottom of the
tank veto  muons by detecting the \v Cerenkov light in water (muon veto system). The veto efficiency is larger than 99.7\% 
for  muon shower events with energy $>$ 4 MeV.\\
A set of analog to digital  (ADC) and time to digital (TDC) converters  records the charge and time information of the PMT pulses for each event.
During the acquisition, a second identical electronic chain is sensitive to the next event
occurring within the following 8.3 ms.
The  electronics can therefore detect pairs of fast time-correlated events.
The coincidence time between the two chains is measured by means of a long range TDC.
Further events are ignored until the first chain is "re-armed" ($\sim$ 20 ms).
For longer delays the computer clock is used providing an accuracy of $\sim$ 50 ms.\\
The trigger condition is set by requiring the signal of 6 PMTs over threshold within a time window of 30 ns.
The corresponding energy threshold is $\sim$ 20 keV with 50\% detection efficiency,
while 99\% efficiency is reached above 90 keV.
The trigger for the second chain is set at a higher value, corresponding to 200 keV
(99\% efficiency).\\
The electronic can be  also triggered by the so-called \emph{afterpulses} which are spurious pulses following genuine PMT output pulse. 
To avoid such effect, the second chain is vetoed for 20 $\mu$s after an event tagged by the muon veto system.\\
The energy response of the detector is calibrated run-by-run by using  the energy spectrum of $^{14}$C decays, naturally present 
in the scintillator.
The measured light yield is $\sim$ 3.6 photoelectrons per PMT for 1 MeV electrons. The electronics saturate at about 6 MeV.\\
The position of the interaction vertex is reconstructed by means of
a maximum likelihood method exploiting the hit time distribution.
The reconstruction algorithm, calibrated by inserting a $^{222}$Rn source in the active volume, provides a resolution of 10 cm at 1 MeV.

\section{Data selection}
\label{sec:select}

The residual cosmic muon flux at  Gran Sasso depth  (3800 m.w.e. maximum depth, 3,200 m.w.e. slant depth) has a  rate
of 1.2 m$^{-2}$ h$^{-1}$ and
an average energy of $\left\langle E_{\mu}\right\rangle =$ 320 $\pm$4$_{stat}$ $\pm$11$_{sys}$ GeV \cite{MACRO}.
The requirements in the  selection of cosmic muons are two-fold: they must be tagged by the muon veto and they must saturate the electronics.
Cosmic muons, crossing the scintillator, produce enough light to blind the detector.\\
For each detected muon, we select the following event in the time window T$_{n}$ = [20, 2000] $\mu$s  as a candidate event for
a neutron capture $\gamma$.
The probability that a random event (R = 0.04 s$^{-1}$ rate) is detected instead of the 2.2 MeV $\gamma$
has an  upper limit equal to T$_{n}$ $\times$ R $\sim$ 8$\times$10$^{-5}$.
We measured the mean capture time of neutrons on protons  equal to 257 $\pm$ 27 $\mu$s, taking into account also events
with double neutron emission.\\
For each  muon-gamma coincidence, $^{11}$C candidates are selected in a subsequent time window T$_w$ = 300 min,
10 times the $^{11}C$ mean life. \\
Random coincidences collected in this window are mainly $^{210}$Bi (Q$_{\beta}$=1.16 MeV) and $^{40}$K
(Q$_{\beta}$=1.32 MeV BR=0.893 and Q$_{EC}$=1.51 MeV  BR=0.107) contamination and external $\gamma$ radiation,
while $^{214}$Bi (Q$_{\beta}$=3.27 MeV) events are discarded through the $^{214}$Bi-Po coincidence.\\
The time profile of the background is expected to be flat on the
scale of 300 minute since the background rate is constant and random coincidences are  not correlated
with cosmic muons. The only bias is introduced by the end of the
data run (typically lasting 2-3 days) which interrupts 8\% of the
selection windows. In such cases the window is completed to 300 min
from a random instant in the run in order to correctly maintain the
background time profile flat.
We estimated that the correspondent probability  to loose a $^{11}$C event is 1\% \cite{Dan05D}.\\
The definition of the optimal energy range of observation, 1.15-2.25 MeV, to detect the $^{11}$C decays,
depends on two main requirements: the enhancement of the signal ($^{11}$C decays)
to background (random events) ratio and the  minimization of the  systematic errors introduced by the energy scale uncertainty.\\
In case  $\gamma$'s from  the positron annihilation  escape the vessel and deposit energy in the water buffer,
the detected energy of the $^{11}$C decay falls below the
observation range. Defining a 0.8 m radius fiducial volume, we reduce non-contained events by a factor 20.
Further, the radial cut avoids distorting optical effects
on the border like the total reflection due to the different refractive indexes of the scintillator and the buffer.\\
The last applied cut exploits the spatial correlation between the $^{11}$C and the neutron capture points.
The events are in fact selected in a sphere of radius \emph{r} centered on the  reconstructed $2.2 MeV \gamma$:
for \emph{r} = 35 cm the background is suppressed by a factor larger than 20 while the signal is reduced
only by a factor $\sim$ 2.\\
The efficiencies and optimal parameters of the cuts here discussed  have been quoted via the Monte Carlo
simulation described in the next section.

\section{The Monte Carlo simulation}

\begin{figure}[t]
\begin{center}
\includegraphics[width=0.95\columnwidth]{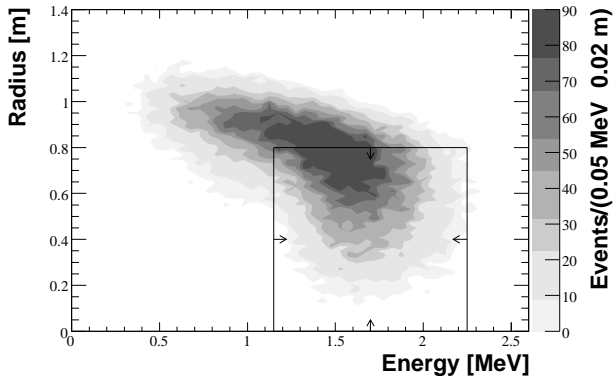}
\caption{\label{fig:scatter} Scatter plot of the simulated $^{11}$C-decay radial position vs energy.
Solid lines represent the  cuts applied in the analysis.
Events with an energy lower than 1.022 MeV are due to the escaping positron annihilation $\gamma$'s.}
\end{center}
\end{figure}
An accurate quantification of the cut efficiencies requires a full simulation
of the $^{11}$C production process from the muon-induced showers originated
in the rock to the  neutron capture and to the  $^{11}$C  decay.\\
The Monte Carlo has been developed in two main steps.
First we  generated and tracked muons and the subsequent cascades
with a FLUKA-based code \cite{Fas01A}.
The code simulates a  320 GeV muon-beam, downward oriented and  uniformly distributed
over the entire CTF water tank.
At this step, the geometry is simplified to only four volumes:
4 m of rocks (CaCO$_3$ and MgCO$_3$)
\cite{Cat86a}, the air, the water  of the CTF tank and finally the scintillator as shown in Figure
\ref{fig:sketch}.  The purpose of the FLUKA-based simulation code is the generation of neutrons in scintillator and
their propagation in the whole detector. \\
In the second step, an \emph{ad hoc} code,  named CTF code \cite{Fra05D,Bor00A},
generates, tracks and reconstructs  $^{11}$C decays  and  $2.2 MeV \gamma$'s  from the neutron capture.
The coordinates of the  neutron production ($\vec{P_p}$) and  capture
($\vec{P_c}$) points from the FLUKA output are input parameters in the CTF code:
$\vec{P_p}$  corresponds to the  origin of the  $^{11}$C-decay while $\vec{P_c}$ is
assumed as the starting position of the $2.2 MeV \gamma$ produced in the neutron capture on hydrogen.\\
The CTF code simulates in detail the detector geometry
including the nylon vessel and the phototubes. Each energy deposit is converted
into optical photons which  are propagated inside the
detector until they are absorbed in the detector material or detected on the PMT's. \\
The tracking code provides a detailed simulation of
the main optical processes like the scintillation light production,
the absorption and reemission processes in the scintillator and
diffusion on the nylon vessel. \\
After all, the same  reconstruction code used in the real data  introduces the energy and spatial resolution effects on the simulated
ones. The final $^{11}$C radial and energy spectra are shown in figure \ref{fig:scatter}.\\
The simulated neutron capture mean time, 254 $\pm$ 1 $\mu$s, is in good agreement with the measured one,
257 $\pm$ 27 $\mu$s.\\
\begin{figure}[th]
\begin{center}
\includegraphics[width=0.95\columnwidth]{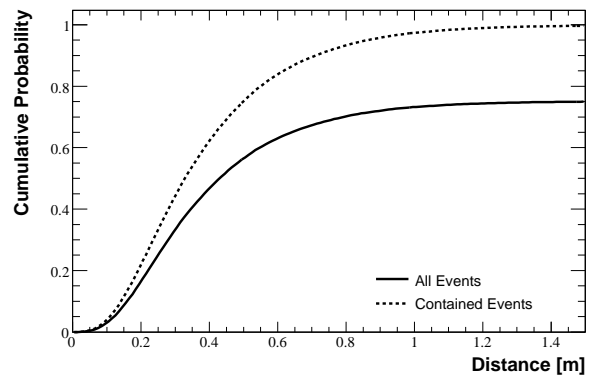}
\caption{\label{fig:cumulative}  Cumulative probability to contain a $^{11}$C event in a sphere of radius r centered on the
 $2.2 MeV \gamma$}
\end{center}
\end{figure}
Secondary particles generating a $^{11}$C event without triggering
the muon veto have been investigated. From the simulation, we expect
mainly $\gamma$'s (91.8\%) and  e$^+$-e$^-$ pairs (8.1\%).
Their  contribution  to  the invisible
$^{11}$C production rate has been estimated in less than 5x10$^{-4}$ day$^{-1}$ (99.99\% C.L.)
by convoluting their rates  with the $^{11}$C production cross sections  \cite{Gal05A}.\\
The main inefficiency in the measurement is due to  neutrons escaping the vessel. If the neutron, indeed, is
captured in water and the subsequent $\gamma$ does not deposit energy in scintillator,
the $\mu-\gamma_{2.2MeV}$ coincidence is not triggered and  the signal is lost.
Neutrons escaping the 1 m CTF vessel account for  26.8\%.
For $\sim$ 50\% of  the fully contained neutrons, the associated $^{11}$C event falls in a 35 cm radius sphere
centered on the reconstructed $2.2 MeV \gamma$, as shown in Figure \ref{fig:cumulative}.\\
All the cut efficiencies are quoted in Table \ref{tab:eff}. \\

\section{The data analysis}

The analyzed  data set corresponds to an effective detector live time of 611 days (June 2002, February 2005). \\
\begin{figure}[th]
\begin{center}
\includegraphics[width=0.95\columnwidth]{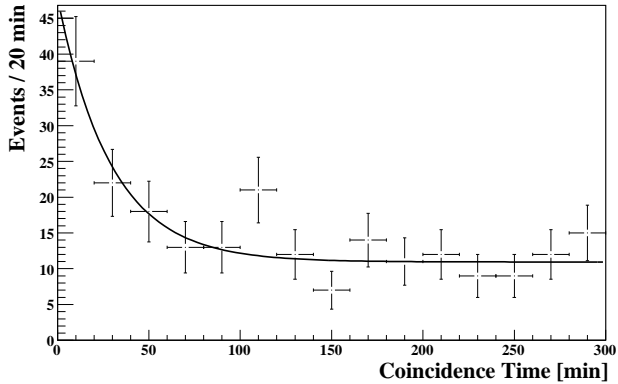}
\caption{\label{fig:c11fit} Fit of the data sample time profile selected by the three-fold coincidence ($\chi^2$/ndf =
9.7/13) with free parameters A and b (see Eq \ref{eq:fit}). }
\end{center}
\end{figure}
The time profile of the data sample selected by the three-fold coincidence technique, shown in Figure \ref{fig:c11fit}, is fitted with:
\begin{equation}
\label{eq:fit}
P(t) = \frac{A}{\tau}e^{-\frac{t}{\tau}} + b ,
\end{equation}
where the free variables in the fit, \emph{A} and $\tau$,
are the number of  $^{11}$C nuclides and the  $^{11}$C  mean life, respectively. 
The fit finds $\tau$ = 27 $\pm$ 11 min (A = 53 $\pm$ 13, b$\times$T$_w$ = 166 $\pm$ 17 and $\chi^2$/d.o.f = 9.7/12) in agreement with the nominal value (29.4 min), 
proving the robustness of the
three-fold coincidence technique.
Moreover, if the 300 min window is started independently from the $\mu-\gamma_{2.2MeV}$ coincidence, the fit is unable to
identify any feature compatible with a decay function.\\
\begin{table}[thb]
\caption{\label{tab:eff} Efficiencies for the $^{11}$C production rate measurement in CTF.}
\vspace*{12pt}
\begin{center}
\begin{tabular}{lll}
\hline
 Efficiency  &  Reason & Value \tabularnewline
\hline
\hline
$\varepsilon_{vis}$  & visible channels & 0.955 \tabularnewline
\hline
$\varepsilon_{end}$ & end of run & 0.990\\
\hline
$\varepsilon_{t}$ &  $\mu$-$2.2 MeV \gamma$ coincidence & 0.925\tabularnewline
\hline
$\varepsilon_{escape}$ & contained neutrons& 0.732\tabularnewline
\hline
$ $ &$^{11}$C energy  E $\in$ [1.15, 2.25] MeV& \tabularnewline
$\varepsilon_{c} $ &$2.2 MeV \gamma$ energy   E $>$ 0.2 MeV & 0.563\tabularnewline
$ $ & $^{11}$C-$2.2 MeV \gamma$ distance   d $<$ 0.35 m&  \tabularnewline
\hline
\hline
Total & &  0.360\\
\hline
\end{tabular}
\end{center}
\end{table}
Performing the fit with $\tau$ fixed to the nominal value, the $^{11}$C production rate is computed from:
\begin{eqnarray}
\label{eq:rate}
\nonumber R(^{11}C) & = & \frac{A}{\frac{4}{3}\pi r^3 \rho  T} \cdot
\frac{1}{  \varepsilon_{vis} \cdot \varepsilon_{end} \cdot \varepsilon_{t} \cdot \varepsilon_{escape}
\cdot \varepsilon_{c}}\\
\\
\nonumber & = & 0.130 \pm 0.026 (stat) \pm 0.014  (syst) day^{-1} ton^{-1}
\end{eqnarray}
(A = 54 $\pm$ 11, b$\times$T$_w$ = 164 $\pm$ 15 and $\chi^2$/d.o.f = 9.7/13) where \emph{r}
is the selected volume radius (0.8 m), $\rho$ the scintillator density (0.88 g/cm$^3$)
and \emph{T} the detector live time (611 days). All the  efficiencies in Eq \ref{eq:rate}  are reported in Table \ref{tab:eff}.\\
The systematic error has been derived by propagating the uncertainties of the reconstruction position ($\sim$ 1.5\%)
and of the light yield ($\sim$ 8.5\%) in Eq \ref{eq:rate}. The systematics takes also into account the stability of the result
when the cut parameters vary around the optimal values.\\
The analysis  measured rate  is in good agreement with the expected one from the CERN experiment:
$0.146 \pm 0.015$  day$^{-1}$ ton$^{-1}$.\\

\section{Discussion}

The success  of the three-fold coincidence technique in selecting $^{11}$C
events and in evaluating correctly their production rate is promising in
prospective of deep underground liquid scintillator detectors.\\ 
The expected rates for \emph{pep}  and CNO neutrinos in Borexino are 
0.021 and 0.035 day$^{-1}$ ton$^{-1}$ (BP2004+LMA+LUNA
\cite{Bah04A,Bah03A,For04A}), respectively. In the energy range of
observation [0.8,1.3] MeV, beyond the $^{7}$Be-$\nu$ electron recoil
energy spectrum, the \emph{pep}+CNO signal, S$_{\nu}$, is reduced  to  
0.015 day$^{-1}$ ton$^{-1}$. In the same window, the expected
contamination  from  $^{11}$C is then about 5 times higher  (B$_{^{11}C}$
= 0.074 $\pm$ 0.008 day$^{-1}$ ton$^{-1}$).\\ 
A second background contribution arises from the trace contaminants in the
scintillator mixture. Assuming for the  $^{238}$U and $^{232}$Th a
concentration level of  10$^{-17}$g/g and 10$^{-15}$g/g for the 
$^{nat}$K, the non-cosmogenic contaminants, B$_{n.c.}$, contribute to the
\emph{pep}+CNO window with 0.006 day$^{-1}$ ton$^{-1}$.\\  
In order to reach a signal-to-background ratio equal to 1, the detection
efficiency of Borexino must be larger than 1 -
S$_{\nu}$/(B$_{^{11}C}$+B$_{n.c.}$) =  0.81. The detection efficiency   is
limited by the physics ($^{12}$C(X,Y)$^{11}$C invisible channels), by the
detector itself (low energy threshold and dead time between sequential
triggers) and by the software cuts in time and space around the neutron
capture $\gamma$'s.  Since, in fact, the three fold coincidence does not identify 
the single $^{11}$C decay but   localizes it in a spherical volume V$_{^{11}C}$, 
the entire   volume V$_{^{11}C}$ must be discarded for a time equivalent to few $^{11}$C lifetimes. Thus,
the main challenge  will  be  the minimization of the detector mass-time
fraction loss.\\  
Assuming a neutron rate of 1.5$\times$10$^{-2}$ $\mu^{-1}$
m$^{-1}$ \cite{Gal05A,Agl99}, we estimate that, even including the trace
contamination,  Borexino can reach a signal-to-background ratio equals to
1, loosing only 14\% of the data \cite{Gal05A,Fra05P}. The optimal cuts
and the relative efficiencies expected for Borexino are quoted in Table
\ref{tab:eff_borex}. \\
Furthermore, the Borexino collaboration is  investigating \cite{Che98P}
the possibility to improve the three-fold coincidence technique by
exploiting the muon track: the reconstruction of the muon track leads, in
fact, to the definition of a cylindrical volume around the track itself.
Intersecting the cylindrical volume with the spherical one centered on the
$2.2 MeV \gamma$, Borexino can efficiently remove $^{11}$C events while
reducing  significantly the fraction  of data loss.\\
\begin{table}[thb]
\caption{\label{tab:eff_borex} Predicted $^{11}$C decay detection efficiencies 
 for the  Borexino detector in order to reach a signal (\emph{pep}+CNO $\nu$s) to 
background ($^{11}C$ decays) ratio equal to 1. The trace contaminanion is assumed at the level of 10$^{-17}$g/g for $^{238}$U and $^{232}$Th and  10$^{-15}$g/g for $^{nat}$K.}
\vspace*{12pt}
\begin{center}
\begin{tabular}{lll}
\hline
 Efficiency  &  Reason & Value \tabularnewline
\hline
\hline
$\varepsilon_{vis}$  & visible channels & 0.955 \tabularnewline
\hline
$\varepsilon_{t}$ &  $\mu$-$2.2 MeV \gamma$ coincidence & 0.989\tabularnewline
\hline
$\varepsilon_{c} $ &$2.2 MeV \gamma$ energy   E $>$ 0.2 MeV & 0.954\tabularnewline
\hline
$\varepsilon_{d} $ & $^{11}$C-$2.2 MeV \gamma$ distance   d $<$ 1 m&  0.984\tabularnewline
\hline
$\varepsilon_{t} $ & $^{11}$C-$2.2 MeV \gamma$ coincidence time   T $<$ 5$\times \tau_{^{11}C}$& 0.993 \tabularnewline
\hline
\hline
Total & &  0.880\\
\hline
\end{tabular}
\end{center}
\end{table}

\section{Concluding remarks}
In this paper we presented the results of  the cosmogenic $^{11}$C measurement based on the three-fold coincidence technique
with the Borexino  Counting Test Facility.  For the first time, deep underground $^{11}$C production
has been detected \emph{in situ}  event by event.\\
The agreement between the measured  $^{11}$C production rate  observed in CTF and the value
extrapolated from the measurement performed at the NA54 CERN facility in a muon on target experiment \cite{Hag00A},
demonstrated that the three-fold coincidence technique is a powerful
tool for isolating and discriminating the $^{11}$C  background.\\
The results also indicate an agreement with the theoretical calculation in \cite{Gal05A}.
When combined with the prediction that the overall
rate of $^{11}$C produced without free neutrons in the final state is limited at 4.5\%,
this observation indicated that Borexino should be
able to minimize the $^{11}$C background at a level compatible with the observation of \emph{pep} neutrinos.\\
In prospective of Borexino, such result opens a new window  in \emph{pep} and CNO neutrino spectroscopy.

\section{Acknowledgements}
We thank D. Motta and E. Resconi  for the useful discussions and comments and  I. Manno, L. Cadonati,  M. Goeger-Neff, 
A. Sonnenschein, A. Di Credico and G. Testera for their past contributions.\\
This work has been supported in part by the Istituto Nazionale di Fisica Nucleare, the Deutsche
Forschungsgemeinschaft (DFG, Sonderforschungsbereich 375), the German
Bundesministerium f\"ur Bildung und Forschung (BMBF), the
Maier-Leibnitz-Laboratorium (Munich), the Virtual Institute for Dark
Matter and Neutrino Physics (VIDMAN, HGF) and the U.S. National
Science Foundation under grants PHY-0201141, PHY-9972127 and PHY-0501118.

\end{document}